\documentclass[aps,twocolumn,floatfix]{revtex4}
\usepackage{times}
\usepackage{graphics}
\usepackage{epsfig,calc}

\usepackage{bm}

\usepackage{amsmath,amssymb,chemarr}

\def\btt1{{\tt$\backslash$\string1}}%

\def\AmS{{\protect\the\textfont2
        A\kern-.1667em\lower.5ex\hbox{M}\kern-.125emS}}

\citeindextrue

\newcommand{\kt}{k_{\text{B}}T}
\newcommand{\pn}{P_\text{N}}
\newcommand{\fc}{P_\text{N}^\text{eq}}
\newcommand{\halftime}{\tau_\text{1/2}}
\newcommand{\gnuc}{g_{\text{nuc}}}
\newcommand{\dg}{\Delta g}

\newcommand{\gc}{g_{\text{c}}}
\newcommand{\gelong}{g_{\text{elong}}}
\newcommand{\nnuc}{n_\text{nuc}}
\newcommand{\nn}{\hat{n}}
\newcommand{\ckt}{c_\text{kt}}
\newcommand{\cc}{c_\text{c}}
\newcommand{\ls}{I_\text{LS}}
\newcommand{\telong}{\tau_\text{elong}}
\newcommand{\tlag}{\tau_\text{lag}}
\newcommand{\tnuc}{\tau_\text{nuc}^\text{max}}

\begin{document}

\title{Understanding the Concentration Dependence of Viral Capsid Assembly Kinetics - the Origin of the Lag Time and Identifying the Critical Nucleus Size}
\author{Michael F. Hagan}
\affiliation{Department of Physics, Brandeis University, Waltham, MA, 02454}
\date{\today}

  \begin{abstract}
The kinetics for the assembly of viral proteins into a population of capsids can be measured in vitro with size exclusion chromatography or dynamic light scattering, but extracting mechanistic information from these studies is challenging. For example, it is not straightforward to determine the critical nucleus size or the elongation time (the time required for a nucleated partial capsid to grow completion). We show that, for two theoretical models of capsid assembly, the critical nucleus size can be determined from the concentration dependence of the assembly reaction half-life and the elongation time is revealed by the length of the lag phase. Furthermore, we find that the system becomes kinetically trapped when nucleation becomes fast compared to elongation.  Implications of this constraint for determining elongation mechanisms from experimental assembly data are discussed.
\end{abstract}

\maketitle

\section{Introduction}
The assembly of protein building blocks into a shell or `capsid' is essential for viral replication and thus understanding the mechanisms by which assembly proceeds could identify targets or opportunities for novel antiviral therapies.  However, despite extraordinary progress in determining the structures of assembled capsids, assembly mechanisms for most viruses remain poorly understood because the structures of transient assembly intermediates have been inaccessible experimentally.  The kinetics for spontaneous capsid assembly in vitro have been measured with size exclusion chromatography (SEC) and dynamic light scattering (e.g. \cite{Prevelige1993,Zlotnick1999,Zlotnick2000,Casini2004,Johnson2005,Chen2008}), but extracting mechanistic information such as the critical nucleus size or the time to assemble an individual capsid has been challenging. In this article, we theoretically examine two models for capsid assembly kinetics to show that these properties can be determined from the concentration dependence of median assembly times and the lag phase.

Assembly kinetics in vitro have been measured for a number of icosahedral viruses (e.g. \cite{Prevelige1993,Zlotnick1999,Zlotnick2000,Casini2004,Johnson2005,Chen2008}) and demonstrate sigmoidal growth characterized by a lag phase, rapid growth, and finally saturation (see Fig.~\ref{fig:timePlots}). Zlotnick and coworkers \cite{Zlotnick1994,Zlotnick1999,Endres2002} showed that partial capsid intermediates assemble during the lag phase, but it has often been assumed that the duration of the lag phase corresponds to the time required for the concentration of critical nuclei to reach steady state, in analogy to models of actin nucleation. However, in this work we show that because light scatter signal measures the mass-averaged molecular weight of assemblages \cite{Casini2004} and SEC usually monitors complete capsids, the length of the lag phase is related to the elongation time, or the time required for a nucleated partial capsid to grow to completion.  Similarly, the critical nucleus size cannot be reliably determined from the concentration dependence of initial or maximum growth rates \cite{Endres2002}, and a method to do so using the extent of assembly \cite{Endres2002} is data-intensive.   We demonstrate that the critical nucleus size can be identified in a straightforward manner from the concentration dependence of the median assembly time. Finally, we show that the system becomes kinetically trapped when the elongation time becomes long compared to the time scale for nucleation.

In a contemporary article related to the current study, Morozov and coworkers \cite{Morozov2009} consider simplified capsid assembly models in which nucleation occurs via a single dimerization event, which enables a elegant analytic solution.  They show that the early phase of assembly can be characterized as a shock front and that for some conditions prohibitively long time scales are required to reach equilibrium.  In this work we consider nucleation as an explicit multistep subunit addition process, with the objectives of understanding the concentration dependence of overall assembly times and learning how nucleation and capsid growth times can be inferred from experimental light scatter measurements.

\section{Models}
Zlotnick and coworkers \cite{Zlotnick1994,Endres2002} have developed a system of rate equations that describe the time evolution of concentrations of empty capsid intermediates
\begin{eqnarray}
\frac{d \rho_1}{d t} &=& -f_1 c_1^2 + b_2 c_2 +\sum_{i=2}^{N}-f_i c_i c_1 + b_i c_i \nonumber \\
\frac{d c_i}{d t}&=&f_{i-1} c_1 c_{i-1} - f_i c_1 c_i \qquad \qquad i=2\dots N \nonumber \\
& & -b_i c_i + b_{i+1} c_{i+1}
\label{eq:ktEmpty}
\end{eqnarray}
where $N$ is the number of subunits in a complete capsid, $c_i$ is the concentration of intermediates with $i$ subunits, $f_i$ is the subunit association rate constant for intermediate $i$, which is related to the dissociation rate constant by detailed balance $b_i=f \exp(\dg_i/\kt)/v_0$, with $\dg_i$ the change in free energy due to association of the subunit, and $v_0$ is the standard state volume. Following Ref.~\cite{Endres2002}, transitions between intermediates are only allowed through binding or unbinding of a single subunit and there is only one intermediate for each size $i$.

We consider two models for intermediate free energies and association rates. In the simple nucleation and growth model (henceforth referred to as `NG')\cite{Zlotnick1999}, the association rate constant $f$ is independent of intermediate size and association free energies are given by $\dg_i=\gnuc$ before nucleation ($i<\nnuc$) and $\dg_i=\gelong$ during elongation ($\nnuc \le i < N-1$), where $\nnuc$ is the critical nucleus size.  Since insertion of the final subunit generates the most new contacts, we set $\dg_{N-1}=2 \gelong$; our conclusions would be the same for an irreversible final assembly step.  Although simple, this model has been shown to reproduce experimental assembly kinetics for several viruses \cite{Zlotnick1999,Johnson2005}. Nucleation is usually assumed to correspond to completion of a polygon (e.g. a pentamer of dimers for CCMV \cite{Zlotnick2000} or a trimer of dimers for turnip crinkle virus \cite{Sorger1986}) and could include intertwining of flexible terminal arms as well as subunit conformation changes. In this work we use $\gelong=2\gnuc$. In Zlotnick and coworkers' formulation, the nucleation and elongation phases are distinguished by having different forward rate constants rather than different association free energies, but the current presentation yields similar behavior and seems easier to justify physically. For simplicity, we neglect geometrical statistical factors ($s_i$ in Ref.~\cite{Endres2002}).

We also consider a model for capsid assembly based on classical nucleation theory (henceforth referred to as "CNT") suggested by the Zandi and coworkers \cite{Zandi2006}, in which the unsatisfied subunit-subunit interactions in a partial capsid intermediate are represented by a line tension $\sigma$, and the binding free energy is
\begin{align}
G_i = i \gc + \sigma l_i  
\label{eq:classNuc}
\end{align}
with $\gc$ the binding free energy per subunit in a complete capsid, $l_i = 2 (\pi/N)^{1/2}[i(N-i)]^{1/2}$ as the number of subunits on the perimeter of the partial capsid and $\sigma = -\gc/2$ \cite{Zandi2006}. The time evolution of intermediate concentrations is solved using Eq.~\ref{eq:ktEmpty}, with the rate constants related by $b_i=f_i \exp[(G_i-G_{i-1})/\kt]/v_0$, and association rate constants are proportional to the perimeter length, $f_i=f l_i$.  This model seems more realistic than the NG model for later stages of capsid assembly, but additional complexity may be required to properly describe nucleation or formation of the first polygon.

We note that both models neglect the possibility of capsids with malformed or non-native structures, which are found for some parameter sets in dynamical simulations \cite{Schwartz1998,Hagan2006,Nguyen2007,Wilber2007,Hicks2006,Elrad2008,Rapaport2008,Nguyen2009}, and association of intermediates \cite{Zhang2006, Hagan2006, Sweeney2008}. We integrate the system of differential equations (Eq.~\ref{eq:ktEmpty}) numerically, with the initial condition $c_1=c_0$, where $c_0$ is the total subunit concentration.  

\section{Estimating time scales for nucleation and elongation}
It is useful to write the overall capsid assembly time $\tau$ as $\tau = \tau_\text{nuc} + \telong$ with $\tau_\text{nuc}$ and $\telong$ the average times for nucleation and growth, respectively.  If we assume the concentration of free subunits is constant during the assembly of a given capsid,
the average time for a capsid to complete the elongation phase in the NG model can be calculated from the mean first passage time for a biased random walk with a reflecting boundary conditions at $\nnuc$ and absorbing boundary conditions at $N$, with forward and reverse hopping rates given by $f c_1$ and $b_\text{elong}=f \exp(\gelong)/v_0$, respectively
\cite{Bar-Haim1998, Hagan2008}
\begin{equation}
\telong=\frac{n_\text{elong}}{f c_1 - b_\text{elong}} - \left(\frac{b_\text{elong}}{f c_1-b_\text{elong}}\right)^2\left(\frac{b_\text{elong}}{f c_1}\right)^{N-n_\text{nuc}}
\label{eq:mfpt}.
\end{equation}
In the limit of $f c_1 \gg b_\text{elong}$ Eq.~\ref{eq:mfpt} can be approximated to give $t_\text{elong}\approx (N-n_\text{nuc})/f c_1$, while similar forward and reverse reaction rates, $f c_1 \approx b_\text{elong}$, gives $t_\text{elong} \approx (N-n_\text{nuc})^2/2 f c_1$. We will see that when elongation is fast compared to nucleation, the duration of the lag phase in capsid completion measurements is given by $\tlag=\telong$ calculated from Eq.~\ref{eq:mfpt} with $c_1=c_0$.

Under conditions of constant free subunit concentration, we could derive the average nucleation time with an equation analogous to Eq.~\ref{eq:mfpt} ~\cite{Hagan2008,Endres2002}
\begin{align}
\tnuc =&\frac{\nn}{f c_0 - b_\text{nuc}} - \left(\frac{b_\text{nuc}}{f c_0-b_\text{nuc}}\right)^2\left(\frac{b_\text{nuc}}{f c_0}\right)^{\nn} \nonumber \\
\approx & f^{-1}\exp\left(G_{\nn} /\kt \right) c_0^{-\nn}
\label{eq:tnuc}.
\end{align}
Because free subunits are depleted by capsid nucleation and growth during spontaneous capsid assembly, however, the nucleation rate never reaches this value and net nucleation asymptotically approaches zero as the concentration of completed capsids approaches its equilibrium value.   Instead, treating the system as a two-state reaction with $\nnuc$-th order kinetics yields an approximation for the median assembly time $\halftime$, the time at which the reaction is 50\% complete
\begin{align}
\halftime = \frac{2^{\nn}-1}{\nn} \frac{\fc}{N f}  \exp\left(G_{\nn}/\kt\right) c_0^{-\nn}
\label{eq:halftime}
\end{align}
with $\nn=\nnuc-1$, $G_{\nn} = (\nn-1)\gnuc$ for the NG model or given by Eq.~\ref{eq:classNuc} for the CNT model, and $\fc$ as the equilibrium fraction of subunits in complete capsids, which can be calculated from the law of mass action \cite{Bruinsma2003, Hagan2006}.  For conditions under which most assembled subunits are in complete capsids, the equilibrium completion fraction is given by \cite{Schoot2007} $\fc/[N(1-\fc)^{N}] = \exp(-G_N/\kt)$ with $G_N$ as the total binding free energy of a complete capsid. The factor of $N^{-1}$ in Eq.~\ref{eq:halftime} accounts for the fact that $N$ subunits are depleted by each assembled capsid.

When capsid growth times are negligible compared to nucleation times, the expressions  Eq.~\ref{eq:mfpt} and Eq.~\ref{eq:halftime} respectively predict the duration of the lag phase and the overall median assembly time.  However, as first noted by Zlotnick \cite{Zlotnick1994} the reaction becomes kinetically trapped if free subunits are depleted before most capsids finish assembling. It was recently suggested \cite{Hagan2008,Morozov2009} that this trap occurs at parameters $\Delta g$ and $c_0$ for which the rate of subunit depletion by nucleation ($N/\tnuc$) is equal to the elongation rate. We find that Eqs.~\ref{eq:mfpt} and ~\ref{eq:halftime} begin to fail a crossover concentration $\cc$ for which initial nucleation and elongation rates are equal, but the system becomes kinetically trapped at a larger concentration $\ckt$ defined by the point at which the median assembly time $\halftime$ matches the elongation time.  These concentrations are related to binding free energies and other parameters by
\begin{align}
 \telong & \approx \tnuc/N &  \quad \mbox{for}\quad c_0=\cc \nonumber \\
\telong &\approx \halftime & \quad \mbox{for} \quad c_0=\ckt
\label{eq:ckt}.
\end{align}
with $\tnuc$ and $\halftime$ respectively given by Eq.~\ref{eq:tnuc} and  Eq.~\ref{eq:halftime}.

\begin{figure} [bt]
\epsfig{file=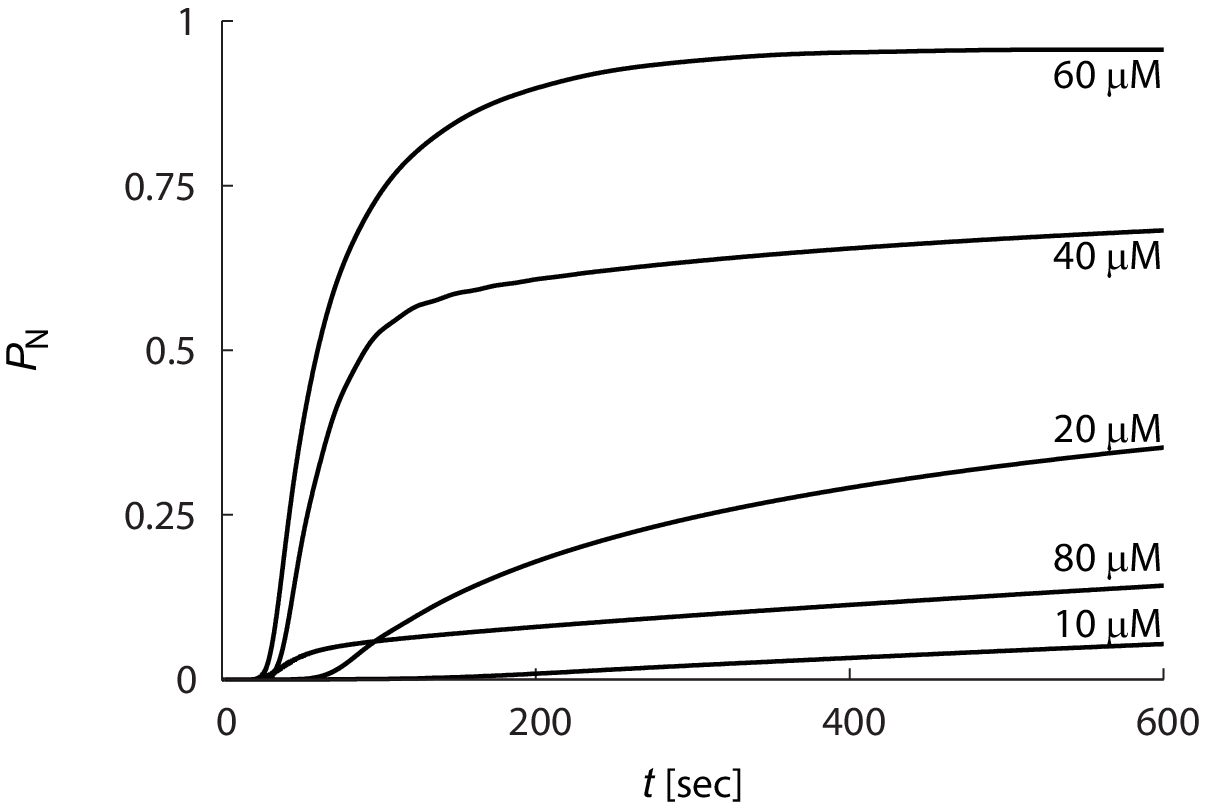,width=.95\columnwidth}
\epsfig{file=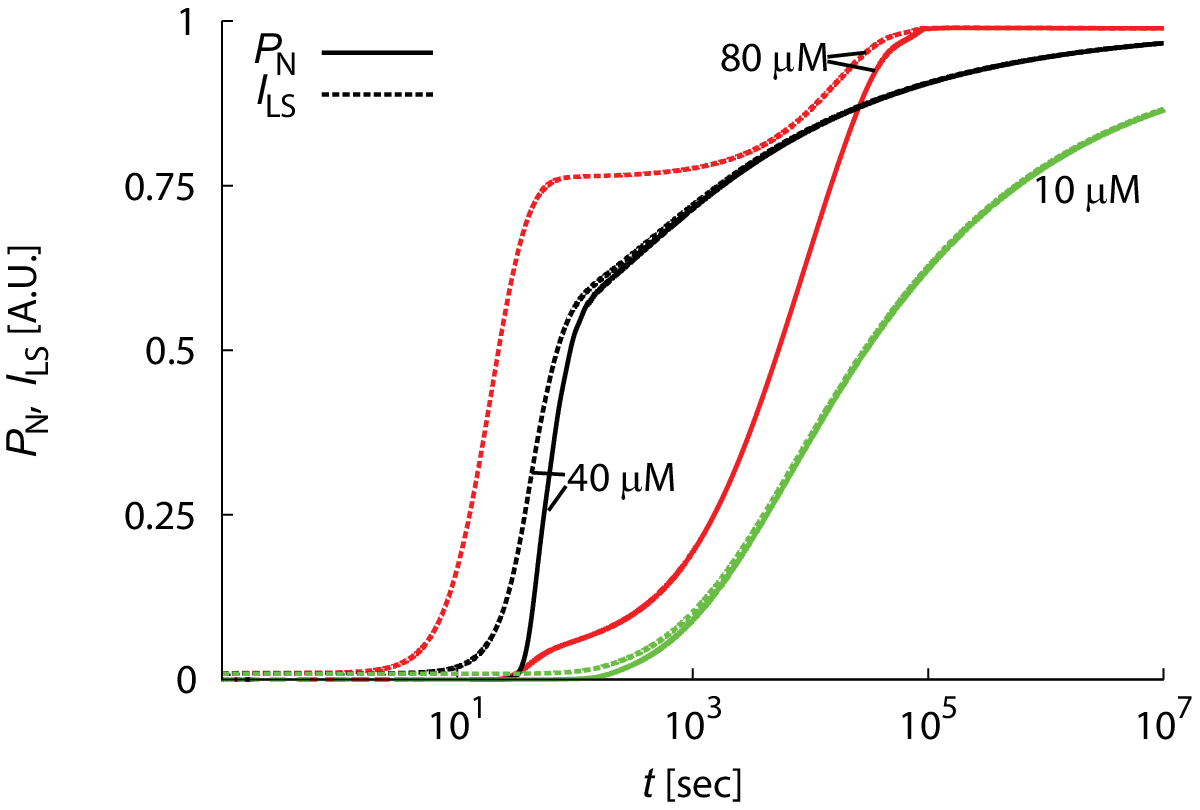,width=.95\columnwidth}
\caption{\label{fig:timePlots}
The time dependence of capsid assembly varies with initial subunit concentration $c_0$. {\bf (a)} The completion fraction $\pn$ as a function of time for indicated initial subunit concentrations.  {\bf (b)} The calculated light scatter closely tracks completion fraction until kinetic trapping sets in. The calculated light scatter (dashed lines) and completion fraction (solid lines) are shown as a functions of time (on a logarithmic scale) for indicated initial subunit concentrations, with $\gnuc=-7 \kt$, $N=120$ and $f=10^5$ M$^{-1}$s$^{-1}$.
}
\end{figure}

\begin{figure} [bt]
\epsfig{file=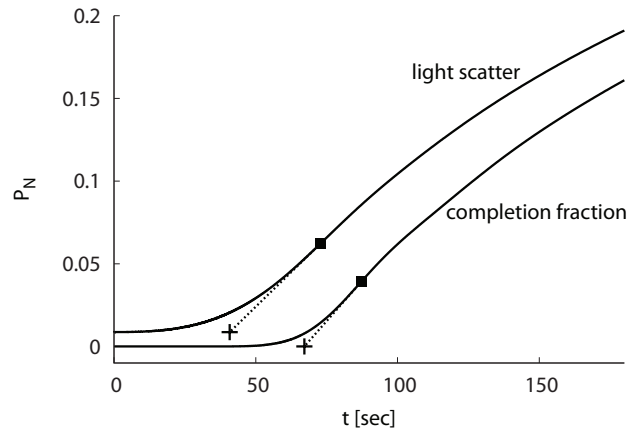,width=.95\columnwidth}
\caption{\label{fig:lagTime}
The end of the lag phase is measured by making a linear fit to the assembly kinetics trace at the point of maximal growth rate ($\blacksquare$).  The lag time (+) then corresponds to time at which the fit (dashed line) crosses the baseline. Plots are shown for $c_0=20 \mu$M.
}
\end{figure}

\section{Numerical results}
We explore the concentration dependence of assembly kinetics by numerically integrating the system of rate equations (Eq.~\ref{eq:ktEmpty}) over a range of initial subunit concentrations $c_0$.  We start with the NG model, with representative parameters: the contact free energy $\gnuc=7 \kt$ ($\approx 4$ kcal/mol) \cite{Ceres2002}, capsid size $N=120$ corresponding to 120 dimer subunits in hepatitis B virus \cite{Ceres2002}, the critical nucleus size $\nnuc=5$, and the subunit association rate constant $f=10^5$ M$^{-1}$s$^{-1}$ \cite{Johnson2005}.	

The fraction of subunits in complete capsids $\pn$, which can be monitored by SEC, is shown as a function of time for several initial subunit concentrations $c_0$ in Fig.~\ref{fig:timePlots}.  In all cases there is a lag time while intermediates assemble, followed by rapid appearance of complete capsids and then eventually saturation of growth. The rate of capsid formation is nonmonotonic with respect to initial subunit concentration; as anticipated by  Eq.~\ref{eq:ckt} a kinetic trap occurs for subunit concentrations larger than $\ckt \approx 60 \mu$M in which the growth of nucleated partial capsids is stymied because the system rapidly becomes starved for free subunits  \cite{Zlotnick1999,Endres2002}.

We calculate light scatter signal $\ls$ from the mass-averaged molecular weight of assemblages \cite{Zlotnick1999}. Since light scatter units are arbitrary, we normalize calculated light scatter by $N$ to give 1 if all subunits are in complete capsids ($\pn=1$).   As shown in Fig.~\ref{fig:timePlots}b, the calculated light scatter closely tracks the completion fraction for initial subunit concentrations below $\cc\approx 38 \mu$M, since the majority of assembled subunits are found in complete capsids once the lag phase is complete. This correspondence completely breaks down by $\ckt$, when there are significant concentrations of partial-capsid intermediates.


\begin{figure}
\epsfig{file=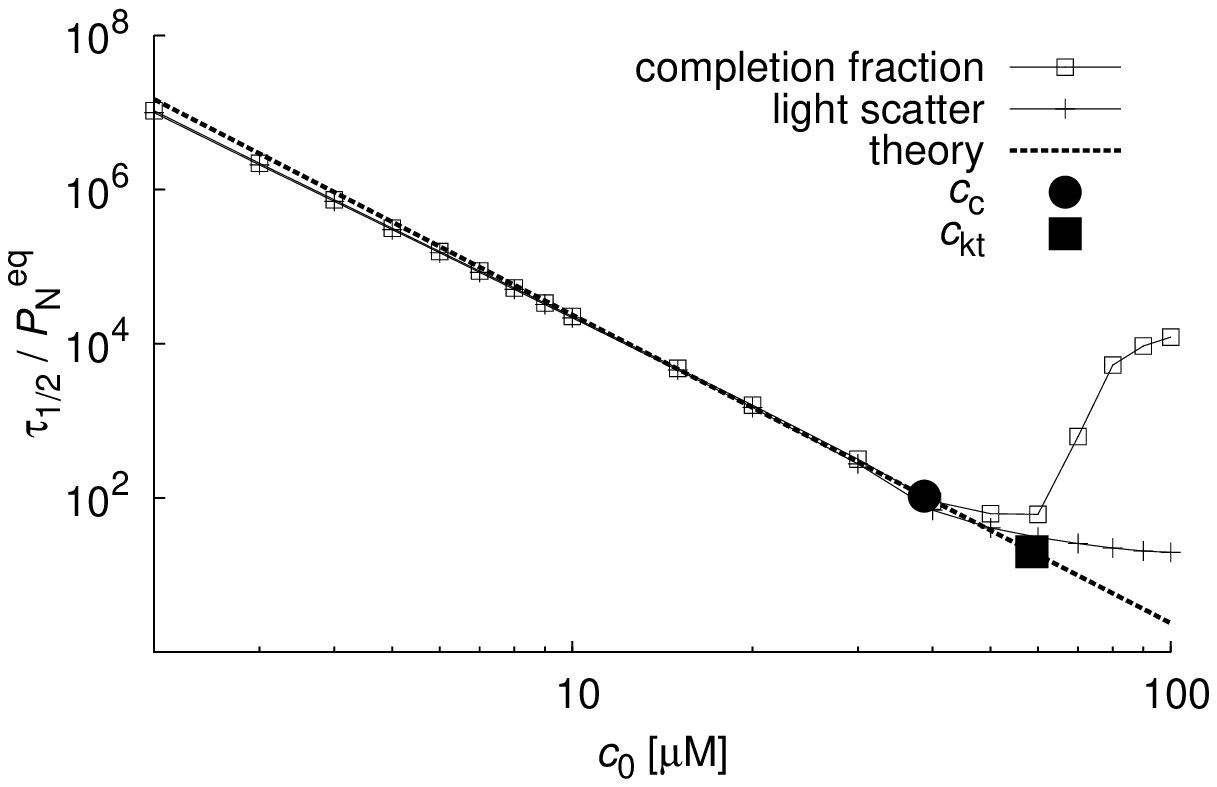,width=.95\columnwidth}
\epsfig{file=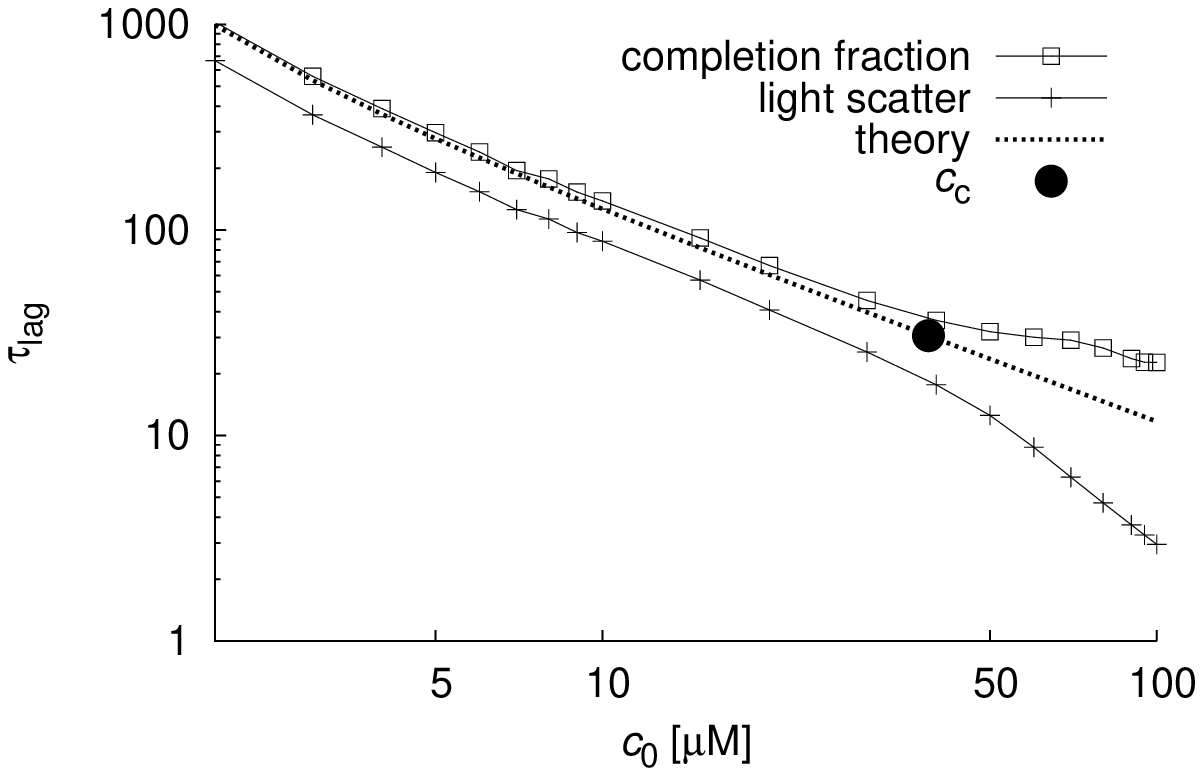,width=.95\columnwidth}
\caption{\label{fig:cScaling}
The median assembly times $\halftime$ {\bf (a)} and the lag times $\tlag$ {\bf (b)} calculated from the completion fraction ($\pn$) and calculated light scatter ($\ls$) are shown as functions of initial subunit concentration $c_0$.  The estimates for the nucleation time  (Eq.~\ref{eq:halftime} with $\nnuc=5$) and lag time  Eq.~\ref{eq:mfpt} are shown as dashed lines, and the estimates for the crossover concentration $\cc$ and kinetic trap concentration $\ckt$ ( Eq.~\ref{eq:ckt}) are shown as symbols on the estimated nucleation curve.
}
\end{figure}

{\bf The concentration-dependence of median assembly times and lag times.}
The median assembly times (reaction half-lives),  and lag times numerically calculated for the nucleation and growth model with respect to the completion fraction $\pn$ and calculated light scatter are given in Fig.~\ref{fig:cScaling}. As illustrated in Fig.~\ref{fig:lagTime}, lag times are extracted from numerical solutions for each parameter set with a linear fit to the assembly trace at the point of maximal assembly rate ($d\pn/dt$ or $d\ls /dt$). The lag time is given by the intersection of the linear fit with the baseline value, which is 0 for the completion fraction and roughly $1/N$ for calculated light scatter.  Note that for initial subunit concentrations below $\cc$, lag times for both the completion fraction and calculated light scatter are inversely proportional to $c_0$, and the mean first passage time estimate (Eq.~\ref{eq:mfpt}) for the capsid elongation phase $\telong$ closely predicts the lag time for completion fractions. The lag time for light scatter is shorter than for the completion fraction because signal is integrated over all of assemblages, but demonstrates the same scaling. Hence, for this model, lag times measured for $\pn$ (SEC) or light scattering are associated with the elongation phase, or the time required to build a complete capsid. This correspondence begins to break down at the crossover concentration $\cc$ (Eq.~\ref{eq:ckt}), when the concentration of free subunits is depleted before the first capsids finish assembling. Above this point, the molecular weight average growth rate becomes dominated by dimerization and hence varies inversely with the square of initial subunit concentration.  The relationship between the lag time and the elongation phase of capsid assembly discovered here explains the observation of Endres and Zlotnick \cite{Endres2002} that the duration of lag time is proportional to the elongation forward rate constant.

The reaction half-life (median assembly time) $\halftime$  is given by $\pn(\halftime )=0.5\fc$ for the completion fraction, or the analogous relation for calculated light scatter. Below $\cc$ half-lives measured with light scatter and completion fraction agree quantitatively, as anticipated from Fig.~\ref{fig:timePlots}b, and agree closely with the two-state nucleation kinetics estimate (Eq.~\ref{eq:halftime}).  In  Fig.~\ref{fig:cScaling}a the numerical and theoretical half-lives are normalized by the equilibrium $\fc$ to emphasize that the scaling with concentration identifies the critical nucleus size: $\ \halftime/\fc \propto c_0^{\nnuc-1}$. The fact that assembly times can be predicted from nucleation kinetics alone can be understood by noting that the elongation time is negligible compared to the overall assembly time below $\cc$. The  close correspondence between the theoretical and numerical median assembly times below $\cc$ suggests that this quantity may provide a simple alternative to the critical nucleus estimator presented in the appendix of Ref.~\cite{Endres2002}.

As anticipated, the kinetic trap point $\ckt$ roughly corresponds to the point at which the time to build the capsid becomes longer than the average nucleation time (Eq.~\ref{eq:ckt}).  Assembly eventually occurs above $\ckt$ as large partial capsids scavenge subunits from smaller intermediates. However, we caution that alternative capsid morphologies \cite{Johnson2005} and/or malformed capsids \cite{Hagan2006,Nguyen2007}, which are not considered in these models, may occur at binding free energies and subunit concentrations above $\ckt$.

As evident from  Eq.~\ref{eq:halftime}, the critical nucleus size can also be identified by evaluating $\ln \halftime/\pn$ as a function of the subunit-subunit association free energy $\gnuc$. We find that agreement between the theoretical predictions and numerical results is insensitive to $\gnuc$ and $\nnuc$.

\begin{figure}
\epsfig{file=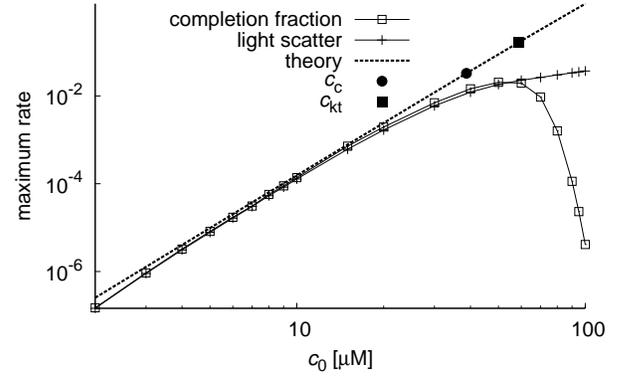,width=.95\columnwidth}
\caption{\label{fig:initRates}
The maximum assembly rates calculated from the completion fraction ($\pn$) and calculated light scatter ($\ls$) are shown as functions of initial subunit concentration $c_0$.  The theoretical estimate for the maximum nucleation rate  (Eq.~\ref{eq:tnuc} with $\nnuc=5$) is shown as a dashed line.
}
\end{figure}

{\bf Maximum assembly rates.} As noted by Endres and Zlotnick \cite{Endres2002}, it is not possible to relate the critical nucleus size to initial assembly rates due to the presence of the lag phase. At low concentrations, though, the maximum assembly rates approach the initial nucleation rate given in Eq.~\ref{eq:tnuc}, as shown in Fig.~\ref{fig:initRates}, and thus nearly scale with $c_0^{\nnuc-1}$.  However, because the maximum assembly rates deviate from the theoretical prediction well below the crossover concentration $\cc$, it appears that median assembly times are a more robust predictor of the critical nucleus size.

\begin{figure}
\epsfig{file=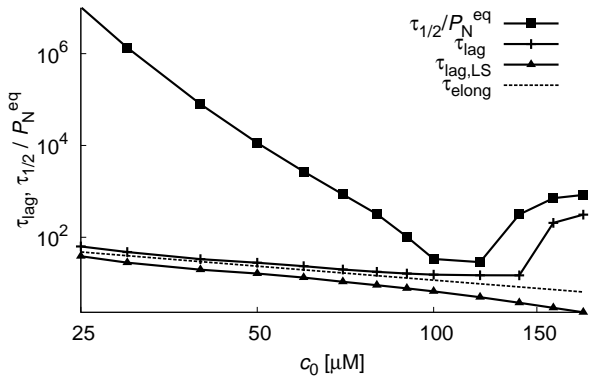,width=.95\columnwidth}
\caption{\label{fig:cNucCscaling}
The median assembly time ($\blacksquare$) and lag time (+) calculated from the completion time and the lag time for light scatter ($\blacktriangle$) for the classical nucleation model are shown as functions of concentration.  The theoretical prediction for lag time  Eq.~\ref{eq:mfpt} is shown as a dashed line.
}
\end{figure}

{\bf The classical nucleation model.}
To evaluate if our conclusions are model dependent, we also consider the concentration dependence of assembly times and lag times for the classical nucleation model (CNT, Eq.~\ref{eq:classNuc}). To facilitate comparison of the two models, we set $\gc=13.8 \kt$  and $f=10^5/\sum_{i=1}^N l_i$ so that $\fc$ and the average association rate constant are the same for both models.  As for the NG model, the calculated light scatter closely tracks the completion fraction below the kinetic trap point, and as shown in Fig.~\ref{fig:cNucCscaling} the lag phase for both quantities lies close to the random walk estimate for the elongation time (Eq.~\ref{eq:mfpt}). The most significant difference between the two models is that the critical nucleus size in the classical nucleation model varies with subunit concentration \cite{Zandi2006}: $\nnuc = 0.5 N (1-\Gamma (\Gamma^2+1)^{1/2}$ with $\Gamma = [\gc - \ln (c_1 v_0)]/\sigma$ with $v_0$ the standard state volume, which gives $9 < \nnuc \lesssim 4$ for the simulated range of initial subunit concentrations.  The reaction half life, however, appears to scale roughly as  $\halftime/\pn \propto c_0^{-9}$ because the effective critical nucleus size increases over the course of the reaction as free subunits are depleted.

For all parameter sets we considered, the CNT model demonstrates a large effective critical nucleus size as the concentration is reduced below $\cc$. Hence, analyzing the concentration dependence of experimental median assembly times could be one way to evaluate which of the CNT model or NG model better represents capsid assembly mechanisms.

\begin{figure}
\epsfig{file=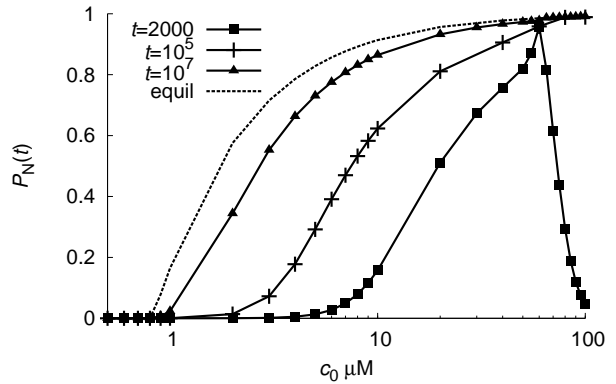,width=.95\columnwidth}
\epsfig{file=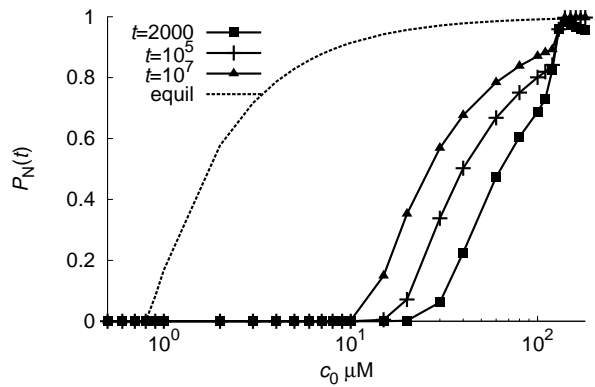,width=.95\columnwidth}
\caption{\label{fig:fc}
Completion fractions at indicated times are shown as functions of initial subunit concentration $c_0$ for {\bf (a)} the nucleation and growth model and {\bf (b)} the classical nucleation model.  In {\bf (a)} and {\bf (b)}, the kinetic trap points $\ckt$  corresponds to the maximum in $\pn$ with respect to $c_0$ at $t=2000$.
}
\end{figure}

{\bf The slow approach to equilibrium.} We note that accurate estimation of $\fc$ is not necessary to obtain the critical nucleus size from $\halftime$, since nucleation rates have such a dramatic concentration dependence even at subunit concentrations for which $\fc>0.5$. This observation could be important, since the median assembly times shown in Figs.~\ref{fig:cScaling}a and \ref{fig:cNucCscaling} extend beyond experimental feasibility at low concentrations (see Fig.~\ref{fig:fc}), particularly for the CNT model because of the large effective critical nucleus size.  Furthermore, as noted by Zlotnick \cite{Zlotnick1994}, capsid assembly approaches equilibrium asymptotically. Therefore, when estimating $\fc$ it is important to plot $\pn(t)$ on a logarithmic scale to judge equilibration, and it may be necessary to extract the contact free energy $\gnuc$ or $\gc$ from fits to kinetic data over a series of concentrations \cite{Zlotnick1999,Johnson2005}. The slow approach to equilibrium due to nucleation barriers was noted from simulations in Ref.~\cite{Hagan2006} and is discussed for the classical nucleation model in Ref.~\cite{Morozov2009}.

\section{Discussion and Outlook}
In this work we examine two theoretical models for capsid assembly, for which we find that the duration of the lag phase measured by SEC or dynamic light scatter is related to the time for a nucleated capsid to grow to completion, and hence scales inversely with initial subunit concentration. When nucleation of new partial capsids is faster than this growth time, the system becomes kinetically trapped due to starvation of free subunits, meaning that capsid formation rates decreases with increasing initial subunit concentration. If there is a well-defined critical nucleus size, it can be identified from the scaling of the median assembly time with respect to initial subunit concentration.  Although highly simplified, the nucleation and growth model we consider here has been shown to closely resemble experimental capsid assembly data \cite{Zlotnick1999,Johnson2005} and our findings therefore suggest new approaches to analyzing and interpreting experimental data to characterize assembly mechanisms.

These predictions have important implications for obtaining mechanistic information about capsid elongation (growth after nucleation) from bulk in vitro assembly kinetics experiments. The fact that overall reaction times are closely predicted by an expression based solely on nucleation kinetics ( Eq.~\ref{eq:halftime},  Fig.~\ref{fig:cScaling}) when assembly is most efficient, suggests that the lag phase contains the most information about elongation.  In fact, the preceding analysis demonstrates that the average elongation time and hence the average growth velocity during elongation are directly related to the duration of the lag phase. Additional mechanistic information about the elongation process could be obtained if the distribution of growth times could be deconvolved from the distribution of nucleation times. The kinetic trap criterion, however, limits this possibility by constraining growth times and hence the lag phase duration to be short compared to nucleation times.

This constraint could potentially be overcome in several ways. The distribution of elongation times can be directly measured in experiments that monitor the assembly of individual capsids, since the elongation and nucleation phases can be separated \cite{Jouvenet2008}.  For bulk assembly studies, recent theoretical studies \cite{Hagan2008,Hagan2009} found that robust assembly is possible under conditions of fast {\bf heterogeneous} nucleation if there is excess capsid protein.  Thus, experimental systems in which capsid assembly is induced by nucleic acids \cite{Johnson2004a}, synthetic polymers \cite{Sikkema2007,Hu2008}, nanoparticles  \cite{Sun2007}, and portal or scaffolding proteins \cite{Parent2005,Parent2006,Tuma2008} could be used to elucidate elongation mechanisms (although assembly mechanisms can be influenced by the heterogeneous component \cite{Johnson2004a,McPherson2005,Hagan2008}).  

Finally, we note that processes not considered in this work, such as transitions between assembly active and assembly inactive conformations of free subunits \cite{Chen2008} or hierarchical assembly \cite{Misra2008} would add additional complexity to analysis of the lag phase. A systematic comparison of model predictions with experimental assembly data over a wide range of concentrations could reveal additional features of complexity in assembly mechanisms and suggest model improvements.

{\bf Acknowledgments} I am grateful to Jane' Kondev for a discussion that led me to investigate the origins of the lag phase. Funding was provided by Award Number R01AI080791 from the National Institute Of Allergy And Infectious Diseases and by the National Science Foundation through the Brandeis Materials Research Science and Engineering Center (MRSEC).


\end{document}